\documentclass[runningheads]{llncs}
\usepackage[T1]{fontenc}
\usepackage{tikz}
\usetikzlibrary{arrows.meta,positioning,fit,backgrounds,calc}
\usepackage{graphicx}
\usepackage{csquotes}
\usepackage{xcolor}
\usepackage[textsize=scriptsize,textwidth=1.9cm]{todonotes}
\definecolor{revcol}{RGB}{0,80,170}   
\definecolor{othcol}{RGB}{0,130,60}   
\newif\ifmarkup \markuptrue           
\markupfalse
\ifmarkup
  \newcommand{\rev}[2]{\todo[color=revcol!15,bordercolor=revcol]{\textbf{#1}}{\color{revcol}#2}}
  \newcommand{\oth}[1]{{\color{othcol}#1}}
  \newcommand{\revmark}[1]{\todo[color=revcol!15,bordercolor=revcol]{\textbf{#1}}}
  \newcommand{\othmark}[1]{\todo[color=othcol!15,bordercolor=othcol]{#1}}
  \newcommand{\delmark}[1]{\todo[color=revcol!15,bordercolor=revcol]{\textbf{#1}: deleted}}
\else
  \newcommand{\rev}[2]{#2}\newcommand{\oth}[1]{#1}
  \newcommand{\revmark}[1]{}\newcommand{\othmark}[1]{}\newcommand{\delmark}[1]{}
  \definecolor{revcol}{RGB}{0,0,0}
  \definecolor{othcol}{RGB}{0,0,0}
\fi

\begin{document}
\title{Constitutive vs. Corrective: A Causal Taxonomy of Human Runtime Involvement in AI Systems}
\titlerunning{A Causal Taxonomy of Human Runtime Involvement in AI Systems}
%
\author{Kevin Baum\inst{1,2,3,4}\orcidID{0000-0002-6893-573X}
\and Johann Laux\inst{2}\orcidID{0000-0003-3043-075X}}

\institute{German Research Center for Artificial Intelligence (DFKI), Saarland Informatics Campus, Saarbrücken, Germany, 
\email{kevin.baum@dfki.de}
\and Oxford Internet Institute, University of Oxford, UK, 
\email{johann.laux@oii.ox.ac.uk}
\and Hamburg University of Technology, Hamburg, Germany
\and Algoright e.V., Saarbrücken, Germany}

\maketitle              
\begin{abstract}
As AI systems permeate high-stakes decision-making, the terminology of human involvement---Human-in-the-Loop (HITL), Human-on-the-Loop (HOTL), and Human Oversight---has become vexingly ambiguous. This complicates interdisciplinary collaboration between computer science, law, philosophy, psychology, and sociology and breeds regulatory uncertainty. We propose a clarification grounded in causal structure, focused on \textit{runtime} involvement. The distinction between HITL and HOTL \rev{R1.13}{is best drawn not spatially---in terms of a human's position \enquote{in} or \enquote{on} a loop---but causally:} HITL is \textit{constitutive} (a human contribution is necessary for the decision output), while HOTL is \textit{corrective} (external to the primary causal chain, capable of preventing or modifying outputs). Within HOTL, we distinguish 
temporal modes---synchronous, asynchronous, and anticipatory---situated in a nested model of provider and deployer runtime. A second, orthogonal dimension captures \textit{cognitive integration}: whether human and machine form complementary or hybrid intelligence, yielding four 
distinct configurations. Finally, we separate these descriptive categories from the normative requirements they serve: statutory \enquote{Human Oversight} is a normative mode of HOTL demanding not merely a corrective causal position but genuine preparedness and capacity for effective intervention. Because the same person may occupy both roles, this role duality must be treated as a design problem requiring architectural and epistemic mitigation.
\end{abstract}
\section{The Tower of Babel of Human Involvement}

\rev{R1.15}{Interdisciplinary discourse about human involvement in AI systems is plagued by semantic misalignment---an ambiguity that the proliferation of literature on human-centric AI has amplified rather than resolved.} \rev{R1.14}{Computer scientists, legal scholars, philosophers, psychologists, and sociologists} frequently use terms such as \enquote{Human-in-the-Loop} (HITL) interchangeably with various modes of human involvement, oversight, or hybrid intelligence. Two examples illustrate the problem. First, in computer science and human-computer interaction, \enquote{human-in-the-loop} describes configurations as diverse as \rev{R1.3}{an annotator whom a learning system repeatedly queries to label the training examples it is currently most uncertain about (so-called active learning; cf.~\cite{MosqueiraRey2023} for a survey of the varied HITL configurations in machine learning)} and a manager who can theoretically override an algorithmic hiring recommendation. These configurations are, as we will argue, distinct in their causal influence on decision outcomes and thus require terminological differentiation. Second, legal discourse in European Union (EU) law frequently equates human oversight with human presence in the decision chain: Article 22 GDPR's requirement that decisions must not be \enquote{based solely on automated processing} is widely read as demanding what we call HITL, yet meaningful human oversight as demanded by Article 14 AI Act is---as we will argue---a structurally different mode of human involvement.

We deliberately decided against conducting a broad literature review. \revmark{R1.1}{\color{revcol}Our contribution is explicative rather than exegetical. Rather than surveying the many competing usages---comprehensive surveys exist for parts of this landscape (e.g., \cite{MosqueiraRey2023})---we put forward a positive proposal and defend it on systematic grounds: conceptual aptness, cross-disciplinary intelligibility, and fruitfulness for operationalization.}\revmark{R1.2}{\footnote{\color{revcol}One strand of literature deserves mention even absent a survey: the HITL/HOTL vocabulary has an especially long history in the debate on autonomous weapons systems \cite{Sharkey2016,Horowitz2015}, where the distinction bears directly on the lawfulness of the use of force. Our taxonomy is meant to generalize and sharpen those earlier formulations.}} We propose that to operationalize human involvement in system architecture and in compliance with regulatory requirements (e.g., Article 14 AI Act), we must move beyond vague metaphors of \enquote*{loops} and instead look at the causal structure of the decision-making process.


Two qualifications apply. First, while our work uses the EU AI Act as background for analyzing notions of human involvement, we do not aim to explicate of the AI Act's own terms. Our aim is broader: a conceptually apt and operationally useful taxonomy, with a focus on AI governance. However, as the first major 
AI regulation, the AI Act 
offers a starting point, and our results should be fruitfully applicable to it. 
Second, our taxonomy is primarily descriptive: it characterizes the causal structure of human involvement in AI systems. We offer these descriptive categories 
because they help formulate and assess normative requirements. Satisfying the descriptive conditions (e.g., being in a HOTL configuration) is typically necessary but not sufficient for meeting normative demands (e.g., providing effective human oversight). Humans must not only occupy the right causal position, but also possess the capacities, information, and authority to make that position meaningful and effective~\cite{sterz2024quest,laux2024institutionalised,baum2026principals,gaube2026keeping}.\othmark{new ref}

\section{The Central Causal Distinction: HITL vs. HOTL}
Before developing our central distinction, we clarify our scope: we focus on \emph{runtime} involvement---a system's operational deployment, in the vocabulary of the EU AI Act---rather than provider-side design choices or post-hoc auditing. However, as we will show (Section~\ref{subsec:hotl}), the boundary between these phases is more permeable than it first seems, especially from the deployer's perspective.

We adopt two role distinctions from the EU AI Act that we deem useful independently of the regulatory context. A \emph{provider} is the entity that develops an AI system or has it developed and places it on the market or puts it into service (cf.\ Art.~3(3) AI Act). A \emph{deployer} is the entity that uses the AI system under its own authority (cf.\ Art.~3(4) AI Act). Each role has an associated temporal perspective (see Figure~\ref{fig:runtimes}). \emph{Provider runtime} spans the entire period in which the system is operational---from initial deployment to decommissioning---during which the provider bears responsibility for the system's design, updates, and life-cycle management. \emph{Deployer runtime} is the operational phase as experienced by the deployer: the period during which the system is actively processing inputs and generating outputs in its specific deployment context. Crucially, the deployer's own cycle of design decisions, operational use, and inspection occurs \emph{within} the provider's runtime (Figure~\ref{fig:runtimes}), a nesting that becomes important for understanding the temporal gradations of HOTL in Section~\ref{sec:hotl-temporal}.

\begin{figure}[t!]
    \centering
    \includegraphics[width=0.65\linewidth]{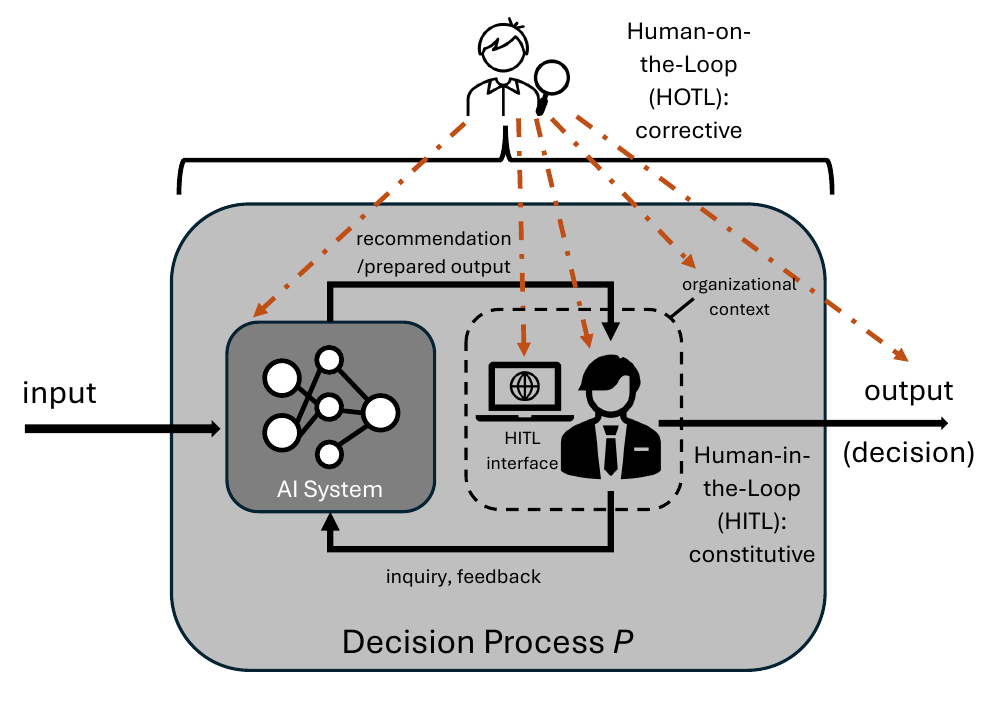}\vspace*{-1em}
    \caption{{\color{revcol}The causal distinction between HITL and HOTL. The Human-in-the-Loop is constitutively embedded within the decision process $P$ (grey box): a necessary link in the default causal chain (solid arrows) from input to output. The Human-on-the-Loop occupies an external, corrective position; dashed arrows mark the possible targets of intervention---the AI system, the HITL's interface, the HITL themselves, the organizational context structuring the process, and the resulting decision---none of which the HOTL is causally necessary for.}}
    \label{fig:hitlhotlstruct}
\end{figure}

Now, why does it matter whether a human is \enquote{in} or \enquote{on} a loop? The spatial metaphor obscures more than it reveals. The influential levels-of-automation framework~\cite{Parasuraman2000} already moves beyond a simple presence/absence dichotomy and embraces a multidimensional view, \rev{R1.4}{but it classifies \emph{degree} rather than kind: for each functional stage, configurations are ordered on a continuum of increasing machine authority. These levels are not causally innocent---each implies something about who acts and who can intervene---yet the ordering passes over the categorical break we are after as just another increment. Between level~5 (the computer \enquote{executes that suggestion if the human approves}) and level~6 (it \enquote{allows the human a restricted time to veto before automatic execution}) \cite{Parasuraman2000}, the human's causal role changes in kind, not merely in degree: an approval requirement is a gate (HITL), a veto window is a switch (HOTL). Our taxonomy is thus complementary to \enquote{Levels of Automation}-style classifications: it names the structural distinction that the scale crosses silently.}
The primary differentiation lies not in proximity but in the nature of a human agent's contribution to the system's output.
We adopt a distinction between \textit{constitutive} and \textit{corrective} human agency \oth{(see Figure~\ref{fig:hitlhotlstruct})}.\footnote{For an early formulation of the constitutive/corrective distinction as regards human involvement in AI systems, cf. \cite{laux2024institutionalised}.}

\subsection{Human-in-the-Loop: Constitutive Agency}

In a HITL configuration, a human is a constitutive element of the causal chain leading to the system's decision\revmark{R1.5}{.\footnote{\color{revcol}
We use \enquote{decision} as a label for the output of $P$, without commitment to a particular output type. The AI Act lists \enquote{predictions, content, recommendations, or decisions} (Art.~3(1); on the definition, see \cite{baum2026aisystem}); our taxonomy is neutral across these---including outputs that trigger actions, whose normative stakes are, if anything, higher. Whether a score, ranking, or recommendation legally counts as a \enquote{decision}---particularly under Art.~22 GDPR---is a separate interpretative question our causal analysis helps frame but does not settle; in \emph{SCHUFA} (Case C-634/21), the CJEU
held that producing a credit score can constitute an automated decision.}}

\begin{definition}[Human-in-the-Loop (HITL)]\label{def:hitl}
A human $H$ is a \textit{human-in-the-loop} with respect to a decision process $P$ if and only if---and if, then because---without $H$'s contribution, no causal chain manifestation of $P$ can effectively generate a decision output. 
\end{definition}

A human action functions as a \enquote*{gate}---a necessary ingredient without which the process cannot proceed. The system pauses or otherwise requires human input to complete its functional execution. This is not merely a matter of being \enquote{involved} or \enquote{consulted}: the human's contribution is an event in the causal chain such that, counterfactually, no output would occur in its absence.

\subsection{Human-on-the-Loop (HOTL): Corrective Agency}\label{subsec:hotl}

In contrast, the Human-on-the-Loop configuration is defined by the autonomy of the primary decision chain. Here, the system can---and by default does---produce outputs without $H$'s contribution. In metaphorical terms: the human is not a \enquote*{gate} but a \enquote*{switch}: present, capable of redirecting the process, but not necessary for it to run.
\revmark{R1.2}{\color{revcol}
We call a causal chain manifestation of $P$ \emph{default} iff it is the chain that unfolds when no agent external to the primary process intervenes---i.e., the path that the system follows in the absence of any corrective human action. This notion is important because a process may admit many possible execution paths, only some of which involve HOTL interventions; the default manifestation is the one that characterizes the system's autonomous behavior.}

\begin{definition}[Human-on-the-Loop (HOTL)]
A human $H$ is a \textit{human-on-the-loop} with respect to a decision process $P$ if and only if---and if, then because---all default causal chain manifestations of $P$ effectively generate a decision output, but there exists a causal chain manifestation of $P$ involving an intervention of $H$ that prevents, modifies, or overrides what would otherwise be the decision output---or $P$ itself.
\end{definition}



\delmark{R1.7} \rev{R1.7}{The human} $H$ is situated externally to the primary causal chain, acting solely in a supervisory or corrective capacity. A human contribution is strictly \textit{corrective}---a veto, a modification, or a confirmation of an otherwise autonomous process. \rev{R1.6}{Such \textit{interventions} can target different elements of $P$ or its output: the AI system, the tooling of an embedded human (where present), that human themselves, the organizational context in which the process runs, or the output before it takes effect (cf.\ Figure~\ref{fig:hitlhotlstruct}).}

\rev{R1.8}{What, then, of a human who monitors and deliberately does \emph{not} intervene? Two cases must be distinguished. If the human's endorsement is causally required for the output to take effect---a mandatory sign-off---the configuration is HITL after all: the endorsement is a gate, however routine (this is what \enquote{effectively} in Definition~\ref{def:hitl} is meant to capture). If it is \textit{not} required, watchful non-intervention makes no causal difference to the token output; the default manifestation runs its course. It is not therefore idle: it certifies the output, alters its modal profile (a defective output \emph{would have been} prevented), and constitutes much of what effective oversight consists in (Section~\ref{sec:ho}; cf.~\cite{sterz2024quest}). Our definitions individuate contributions causally at the level of token outputs; the epistemic and normative significance of vigilant non-intervention belongs to the normative layer the taxonomy serves.}
Note, however, that when a HOTL does intervene and change the decisional direction, they become causally necessary for the specific outcome that results from the intervention.\footnote{
More precisely: call $H$ an \textit{outcome-HITL} with respect to $P$ and a specific output $o$ iff $o$ was produced by a causal chain manifestation of $P$ in which $H$'s contribution was a necessary part. The main-text definitions, by contrast, characterize \textit{process}-level roles, defined over all (default) manifestations of $P$. Process-HITL implies outcome-HITL for every output of $P$; process-HOTL becomes outcome-HITL exactly for those outputs shaped by an intervention. This explains why (synchronous, see below) HOTL can \enquote*{phenomenologically resemble} HITL: upon intervention, both occupy the same causal role with respect to the token output; the difference lies in the counterfactual structure across possible outputs---HITL is necessary for all, HOTL only for those actively shaped.\label{fn:outcome-hitl}
}

\subsection{Temporal Gradations within HOTL}\label{sec:hotl-temporal}

Not all HOTL configurations are equal. The temporal relationship between a human's corrective capacity and the decision process matters significantly for whether HOTL can satisfy normative requirements such as effective oversight. We distinguish three
temporal modes, the first two of which share a common logic, while the third is structurally different.

\paragraph{Synchronous HOTL:} A human monitors operations in real-time and can intervene immediately. This is the classical \enquote{human on the loop} scenario---a human watching a dashboard with capacity for immediate intervention. 
\delmark{R1.9}

\begin{figure}[t!]
    \centering
    \includegraphics[width=0.75\linewidth]{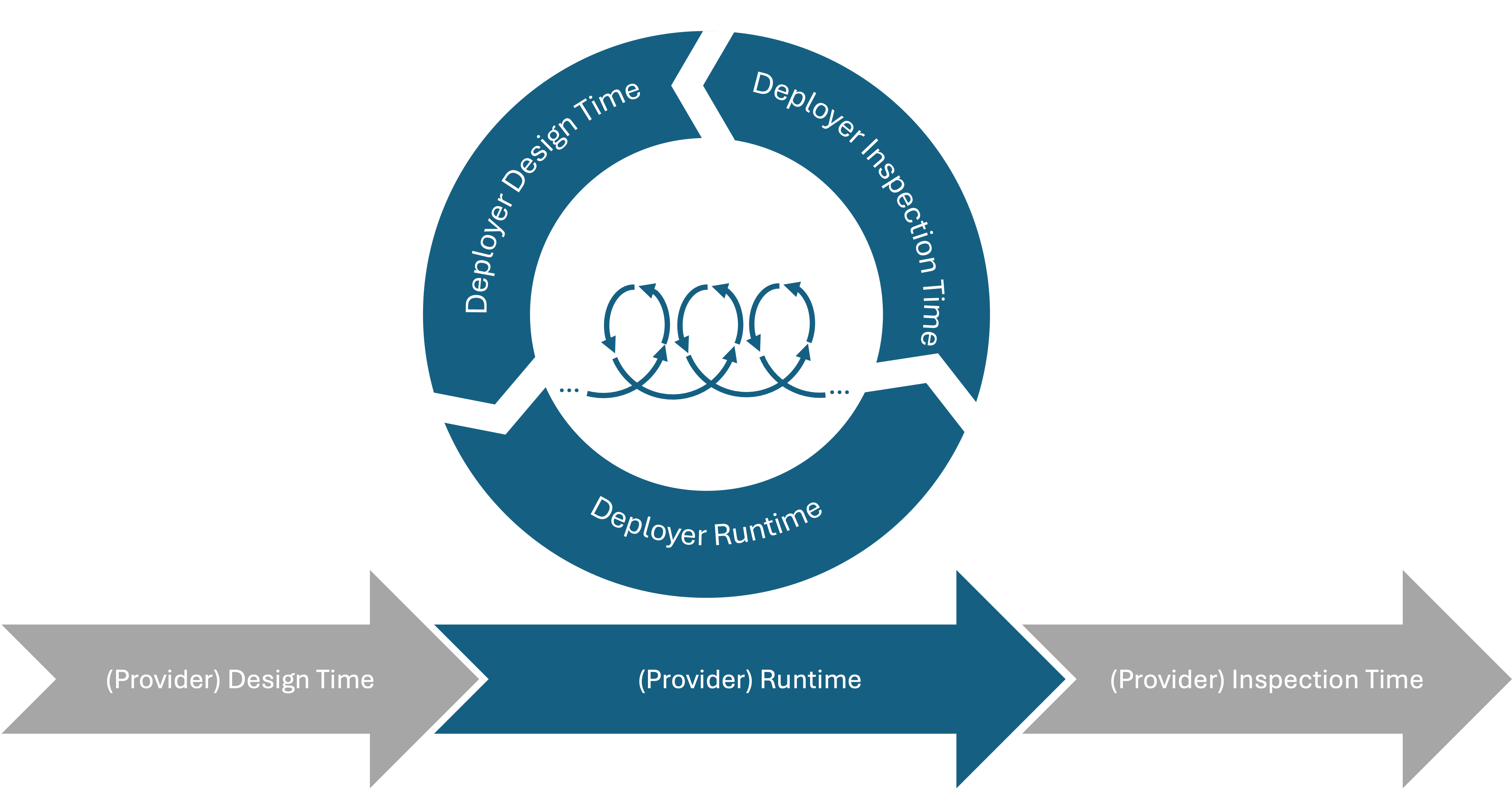}
    \caption{
Nested temporal perspectives on AI system life-cycles. The deployer's design-runtime-inspection cycle operates within the provider's runtime phase. Human oversight interventions may target different phases of this nested structure: synchronous HOTL intervenes during deployer runtime; asynchronous and anticipatory HOTL shape deployer design time based on inspection-time findings.
    }
    \label{fig:runtimes}
\end{figure}

\paragraph{Asynchronous HOTL:} A human reviews system operations periodically (e.g., daily logs, weekly audits) and can intervene with some delay. The corrective capacity exists, but is temporally displaced from individual decisions: the intervention targets future system behavior (e.g., by adjusting parameters or triggering retraining) rather than the decision currently being made. In the terms of Figure~\ref{fig:runtimes}, asynchronous HOTL operates during deployer \emph{inspection time}, shaping the next iteration of deployer \emph{design time}---thus falling outside deployer runtime but within provider runtime.
\oth{Where periodic review occurs before outputs take effect (e.g., decisions released only after batch review), the configuration is, in our terms, synchronous with latency: the distinguishing feature is not elapsed time but whether intervention can occur before the token output becomes effective.}

\medskip
Synchronous and asynchronous HOTL share a reactive logic: the human observes system behavior---whether concurrently or retrospectively---and corrects whenever intervention is warranted. They differ in latency, but not in kind. A third mode is structurally different:

\paragraph{Anticipatory HOTL:} A human exercises corrective influence prospectively by establishing normative constraints, authorization boundaries, or alignment parameters before the system operates. A human is not monitoring at deployer runtime (but still at provider runtime). Instead, the human has shaped the space of permissible actions in advance. This mode connects to work on indirect human oversight through personalized AI alignment, where justifiability and verification of the system's \enquote{understanding} of normative expectations become central~\cite{baum2026principals}\delmark{R1.9}.
\footnote{
What distinguishes anticipatory HOTL from ordinary system design, given that every design choice constrains future behavior? Three features jointly: it occurs within provider runtime, targeting a live decision process; it is oriented toward a specific deployment context rather than system behavior in general; and, most crucially, it is subject to iterative refinement---the human can inspect operational behavior and revise the normative constraints accordingly (cf.\ the deployer's design-runtime-inspection cycle in Figure~\ref{fig:runtimes}). A choice that is not revisitable within the system's life-cycle is design, not corrective involvement. Whether an instance of anticipatory HOTL also qualifies as Human Oversight depends on Definition~\ref{def:ho}.\label{fn:anticipatory-design}
\label{fn:anticipatory-design}}

\medskip
With all three modes now in view, we can note a further structural distinction. Only synchronous HOTL can become outcome-HITL for a given deployer-runtime decision (cf.\ footnote~\ref{fn:outcome-hitl}):
upon intervention, the human becomes causally necessary for the specific output. Neither asynchronous nor anticipatory HOTL can achieve this, because their corrective influence operates at the level of the decision \emph{process}, not the individual decision \emph{token}: asynchronous HOTL intervenes after the decision has been made, shaping future iterations; anticipatory HOTL shapes the decision space before deployment.

Indeed, anticipatory HOTL is not merely a temporal displacement of reactive oversight but a different \emph{kind} of control and accountability structure: it operates on the conditions under which future decisions are made rather than on individual decisions or their outputs (cf. \cite{baum2026principals}). Where synchronous and asynchronous HOTL ask \enquote{was this decision acceptable?}, anticipatory HOTL asks \enquote{is the space of permissible decisions adequately constrained?} This structural difference has significant implications for agentic AI systems, where the speed, opacity, and scale of autonomous action may render reactive oversight modes insufficient while anticipatory oversight remains viable. A fuller development of anticipatory human oversight as a governance paradigm for agentic AI is beyond the scope of this taxonomy but is currently being pursued by the authors.

These gradations highlight that HOTL encompasses configurations with different capacities for effective intervention. Importantly, while asynchronous and anticipatory HOTL intervene outside of \textit{deployer} runtime, they remain within \textit{provider} runtime and oriented toward the runtime decision process (Figure~\ref{fig:runtimes}).

\medskip
\noindent\textbf{In short:} HITL is a necessary link in the causal decision
chain; HOTL is external control over a chain \oth{that otherwise---i.e.,
absent intervention---runs independently of that human.}
\medskip

\noindent\textbf{Boundary Cases and Process Individuation.}\label{subsubsec:boundary}
%
%
The categorization of a system as HITL or HOTL depends on how we individuate the decision process $P$. Consider an autonomous vehicle that requires human confirmation for lane changes (HITL with respect to lane-change decisions) but can autonomously select routes to avoid traffic (HOTL with respect to routing---a human can override, but the system proceeds by default). The same vehicle executes emergency braking autonomously; here, the human is not meaningfully a synchronous HOTL: the intervention happens too fast for real-time correction. However, if a human can adjust parameters governing emergency braking behavior---such as sensitivity thresholds or trigger distances, perhaps within a legally mandated range---this constitutes anticipatory HOTL. The human shapes the space of permissible system behavior in advance, even though they cannot intervene in the moment of emergency. This illustrates that the same person may simultaneously occupy HITL, synchronous HOTL, and anticipatory HOTL roles with respect to different subsystems or phases of a complex system---\oth{an instance of the role duality (here, in fact, \textit{plurality}) discussed} further in Section~\ref{sec:duality}.

\section{Cognitive Integration: Complementary vs. Hybrid} 

While the distinction between HITL and HOTL describes a human's \textit{causal} role, a second dimension is necessary to describe the \textit{cognitive} relationship between human and machine. 
\rev{R1.10}{It is needed because causal position underdetermines cognition: two humans occupying identical causal roles may differ fundamentally in whether their judgment is formed independently of the system or through it. Since the value of human involvement---for instance, as an independent check---often turns on precisely this difference, a taxonomy should capture it.}
We propose distinguishing between \textit{Complementary Intelligence} and \textit{Hybrid Intelligence}.

\subsection{Complementary Intelligence}

In complementary settings, the human and the AI system \rev{R1.11}{do not form an integrated cognitive system as specified below}. Instead, they work side-by-side, each assuming tasks according to 
comparative advantages.

The contributions of the human and the AI system are functionally separated but mutually supportive; they complement each other without forming an inseparable decision-making entity. Each party's formation of decisions remains autonomous, and the overall output arises through aggregation or coordination, not through a shared cognitive system.
Take breast cancer screening as an example. The AI system scans mammograms and flags likely abnormalities. The radiologist provides clinical judgment by reviewing the image, considering the patient's history, and deciding whether to recommend follow-up tests. 


\subsection{Hybrid Intelligence}
\revmark{R2.1, R1.11}{\color{revcol}
Hybrid intelligence arises when human and machine form an \emph{integrated cognitive system}.\footnote{We use \enquote{integrated cognitive system} in the functional sense familiar from the extended-cognition literature
\cite{clark1998extended,heersmink2015dimensions}, without commitment to stronger claims about, e.g., phenomenal unity or joint agency. Where we speak of cognitive \enquote{entanglement} below, this is an informal gloss on the same relation.} The mark of hybridity is \textit{representational interdependence}: the human's task-relevant cognitive states (beliefs, judgments, situational awareness) are partially constituted by or dependent on the machine's representations---and, insofar as the system adapts to its user, vice versa. A human thinks \textit{through} the system rather than merely \textit{with} it. By contrast, in complementary configurations, each party maintains autonomous representational states that are then coordinated.
Crucially, what is jointly constituted in hybrid configurations is the \emph{judgment}, not necessarily the \emph{output}: the human's judgment could not have the same content outside the coupling. Whether that judgment is causally necessary for the system's output (HITL) or merely corrective (HOTL) is a further, independent question---which is precisely what makes the cognitive dimension orthogonal to the causal one.} 

\oth{Augmented reality (AR) overlay surgery illustrates the idea: the surgeon perceives internal structures beneath tissue and bone through 3D digital imaging and operates through this AR-mediated perception; we return
to this case below.} 

The operative test for distinguishing complementary from hybrid intelligence is counterfactual: in complementary configurations, the human could in principle form the same judgment through independent means---the system adds information but does not alter the structure of the human's reasoning. In hybrid configurations, the human's judgment could not have the same content without the system's representations; the system partially constitutes what the human perceives or how they reason, not merely what data is available to them. This distinction admits of degrees rather than a sharp boundary---a surgeon whose perception is restructured by an AR overlay is a clearer case than an analyst whose intuitions have been shaped by long exposure to a predictive model. Operationalizing this spectrum is an empirical question connecting to work on extended cognition~\cite{clark1998extended}, cognitive integration~\cite{heersmink2015dimensions}, and hybrid intelligence~\cite{dellermann2019hybrid}.

\rev{R1.12}{Complementarity and interdependence are not opposites. In the human--AI teaming literature, teams are characterized by both: members contribute complementary knowledge, skills, and abilities \emph{and} are interdependent regarding a shared task \cite{oneill2022teaming,seeber2020machines}. Our distinction cuts along a different line: not \emph{task} interdependence (mutual reliance for 
an outcome) but \emph{representational} interdependence (the partial constitution of one party's cognitive states by the other's representations). A human--AI team whose members rely on each other's contributions while each forms their judgment autonomously instantiates complementary intelligence, 
however tightly coupled the task structure; hybridity begins only where the judgment itself is constituted through the system's representations.}


\subsection{Interaction of Dimensions}

The causal dimension (HITL/HOTL) and the cognitive dimension (complementary/hybrid) are conceptually orthogonal, yielding four possible configurations:

\paragraph{HITL + Complementary:} A human is constitutively necessary but maintains a cognitive process that is separate from the system. Consider a judge using a recidivism risk assessment tool in a criminal proceeding: the system produces a risk score, but the judge must still decide whether parole is granted or which sentence is appropriate in the case at hand. The judge weighs the score alongside independent legal reasoning, case-specific knowledge and impressions, and normative judgment---two functionally autonomous cognitive processes coordinated at the point of decision, but neither constituting the other. This configuration is common in decision-support systems.

\paragraph{HITL + Hybrid:} A human is constitutively necessary and thinks through the system's representations. Consider a surgeon operating with an augmented-reality robotic surgery system that fuses pre-operative imaging, real-time sensor data, and AI-generated tissue classifications into the surgeon's visual field. The surgeon must act for the procedure to proceed, and their perceptual and decisional states are partially constituted by the system's representations---they perceive tissue boundaries and risk zones that would not be cognitively available without the overlay. The same surgical judgment could not be formed outside the integrated cognitive \rev{R1.11}{system}.

\paragraph{HOTL + Complementary:} A human monitors an autonomous system and intervenes based on independent judgment. Consider a safety operator overseeing an autonomous warehouse robot fleet via a dashboard: the robots navigate and pick items autonomously, while the operator watches aggregate metrics and camera feeds, intervening when their own assessment of warehouse logistics warrants it. The operator's cognitive process remains separate from the robots' planning---a supervisory role exercised from both cognitive and causal distance.

\paragraph{HOTL + Hybrid:} Consider a clinician overseeing an AI-driven ICU early-warning system. The system autonomously monitors patients and generates alerts; the clinician occupies the corrective position---able to override triage priorities, adjust alarm thresholds, or escalate a case to senior staff or \oth{the system provider} when the system misses a pattern. What makes this hybrid is that through sustained interaction across the deployer's design-runtime-inspection cycle \oth{(cf.\ Figure~\ref{fig:runtimes})}, the clinician's understanding of dangerous patient trajectories has been reshaped by the system's predictive models. When they exercise corrective judgment---deciding whether to act on an alert or intervene where the system has not---they draw on a cognitive framework partially constituted by the system's representations. They do not merely receive outputs and assess them independently (that would be complementary); they have internalized the system's way of perceiving risk, and this reshaping persists even when they override the system. The cognitive entanglement here is built up iteratively through the inspection-time spiral rather than occurring concurrently during runtime---a temporally displaced form of hybrid intelligence. 
\revmark{R2.1}{\color{revcol}
Note what is and is not jointly constituted here. The default outputs---the alerts---are generated autonomously; that is exactly what makes the configuration HOTL rather than HITL. What is hybrid is the \emph{corrective judgment}: the machine cannot produce it, and the clinician could not form a judgment with that content without the internalized representations. Hybridity attaches to the oversight function, not to the primary output.} \rev{R2.1}{This also makes the configuration \textit{normatively} precarious: effective oversight presupposes a capacity for assessment at least partially independent of the system under scrutiny---precisely what representational interdependence erodes (cf.~\cite{laux2025automation} and Section~\ref{sec:duality}).}

\medskip
In practice, hybrid intelligence is 
commonly associated with HITL configurations, \revmark{R2.1}{\color{revcol}since the sustained, engaged interaction from which cognitive entanglement arises is most naturally afforded by constitutive involvement in the decision process}. The HOTL + Hybrid case illustrates that cognitive entanglement does not \textit{require} constitutive involvement---it merely tends to accompany it.

Bainbridge's~\cite{bainbridge1983ironies} seminal analysis of the \enquote{ironies of automation} is instructive here: automation does not eliminate human involvement but transforms it, often making the residual human role more cognitively demanding precisely when a human has less practice and engagement. Moving from HITL to HOTL does not reduce cognitive demands; it changes them in ways that may make effective oversight \textit{more} difficult. The HOTL role requires sustained vigilance without the engagement that comes from constitutive participation---a tension particularly acute in complementary configurations where humans must maintain independent situational awareness of a process they are not actively shaping.

\section{The Nature of Human Oversight (HO)}\label{sec:ho}

Finally, we address a frequent confusion regarding \enquote{Human Oversight}, particularly under the EU AI Act. Conceptual clarity requires understanding Human Oversight as a \textit{normative} mode of HOTL.

\subsection{Oversight Implies HOTL}

Human oversight designates functions oriented toward supervision, monitoring, and intervention vis-\`a-vis a running (potentially autonomous) system process. By definition, this role is external to the primary causal decision chain---it is corrective, not constitutive. This structural claim has a counterintuitive but important regulatory implication. Article~22 GDPR's requirement that decisions not be \enquote{based solely on automated processing} is naturally read as demanding HITL: a human must be part of the decision process. Yet a merely formal HITL configuration---in which the human does little more than rubberstamping otherwise automated outputs---provides no meaningful oversight. Article~14 AI Act, by contrast, requires that high-risk AI systems be \enquote{effectively overseen}. On our account, this formulation demands genuine HOTL capacity: the human overseer must be systematically prepared and positioned to monitor and intervene. A well-implemented HOTL configuration under Article~14 AI Act may thus provide \emph{stronger} protection than the HITL configuration that the wording in Article~22 GDPR prima facie seems to demand (see also footnote~\ref{fn:gdpr-aia}).\footnote{\label{fn:gdpr-aia}Our taxonomy helps explain why Art.~22 GDPR's negative framing is regulatorily insufficient, and why Art.~14 AI Act's affirmative \enquote{effective oversight} standard 
requires more than mere HITL presence. On its wording, Art.~22 GDPR asks whether a human was \emph{present} in the decision chain; Art.~14 AI Act asks whether a human was \emph{effective} as an overseer. The former can be satisfied by a perfunctory gate; the latter requires a genuine switch with the capacity and preparedness to be pulled.}

\begin{definition}[Human Overseer (HO)]
A human $H$ is a \textit{human overseer} with respect to a decision process $P$ if and only if---and if, then because---$H$ is systematically prepared for and is in a position to consciously monitor operations and intervene, if necessary, in order to substantially reduce AI-induced risks.\footnote{Our definition of human oversight shares structural features with the 'meaningful human control' framework of~\cite{SantoniDeSio2018}, which grounds control in reason-responsiveness via tracking and tracing conditions. However, our taxonomy is deliberately more minimal. We specify causal positioning and preparedness without committing to a particular theory of moral responsibility (but see \cite{Fischer1998,baum2022responsibility,sterz2024quest}). The meaningful human control literature provides a natural complement for operationalizing the normative dimension we flag but do not fully develop here.}\label{def:ho}
\end{definition}

The qualifier \enquote{substantially} is not mere hedging. Operationalizing what counts as sufficient risk reduction is a genuinely difficult problem that we address elsewhere~\cite{langer2025complexities,gaube2026keeping}.\othmark{new ref} The challenge lies in specifying effectiveness thresholds when there is no ground truth for \enquote{safe enough} and when the effectiveness of oversight depends on a complex interplay of technical, individual, and contextual factors.

In other words, being a human overseer is a specific HOTL role---one that adds normative requirements (systematic preparation, genuine capacity, risk-reduction function) to the descriptive HOTL category. Thus:

\medskip 
\noindent\textbf{Lemma:} Human Oversight implies a HOTL role: qua overseer, a human is \textit{not} a necessary link in the default decision chain and \textit{can} intervene without being involved in every individual decision.

\medskip 

This has an important implication: a system can be descriptively HITL while failing to provide human oversight in the normatively relevant sense. A human who must click \enquote{approve} for every decision (HITL) but rubberstamps without genuine judgment does not thereby become an overseer.\footnote{Empirical evidence supports this concern: Green and Chen~\cite{Green2019} demonstrate that providing human decision-makers with algorithmic risk assessments can systematically alter---and in some cases worsen---the quality and fairness of their decisions, even when those humans occupy a formally constitutive position in the decision chain.} Conversely, effective HOTL with genuine monitoring capacity may provide better oversight than perfunctory HITL. Our causal taxonomy clarifies structure; it does not guarantee that normative requirements are met.

\subsection{Role Duality vs. Role Identity}
\label{sec:duality}

A common counter-argument involves systems where oversight is mandated, yet humans are also involved in the loop. 
\oth{Nothing---conceptually or logically---forbids assigning the safety
function of Human Oversight to a person who also occupies a HITL role in the
operative process itself, e.g., for input validation, action authorization,
or calibration.}
This is consistent with our proposal, but it underscores the need for a strict separation of roles. 
Consider again the autonomous vehicle example. 
\oth{During normal operation, the driver may be HITL with respect to destination-level navigation (the system cannot set or change the planned destination without human input) while simultaneously being HOTL with respect to the autonomous driving functions---including the dynamic
rerouting discussed in Section~\ref{subsubsec:boundary}---with the capacity to take over if the
system malfunctions.}

While the same physical person may hold both positions, the roles remain conceptually distinct: as \textbf{HITL}, the person is a constitutive part of the decision chain; as \textbf{HOTL}/overseer, they monitor the system from outside it.
%
\oth{Uniting both functions in one person} does not dissolve the conceptual boundary: HITL remains constitutive, while Human Oversight remains a specific, normative form of corrective HOTL.
This role duality creates genuine tensions. If a human is required to act for the process to function (HITL), they are an \enquote{actor} within the process, a component of the machine's operational logic, rather than an overseer of it. One can oversee a process that one runs, but this configuration introduces role conflicts, conflicts of interest, and the 
danger of automation bias and other biases~\cite{laux2025automation}. The HITL role creates engagement and familiarity, but may compromise the critical distance that effective oversight requires.

However, sometimes the only plausible overseer is the person already in the HITL role, and introducing another human as genuine HOTL may be infeasible. Two responses are available. The first is architectural: layered oversight structures---in which
real-time oversight (monitoring individual decisions), systemic oversight (reviewing aggregate patterns and system behavior), and compliance oversight (verifying adherence to legal and organizational requirements)  are distributed across different roles and organizational levels---can compensate for compromised independence at the real-time layer. When the frontline operator necessarily occupies both HITL and HOTL roles, a structurally independent overseer at the systemic or compliance layer can provide the critical distance that the frontline role duality undermines \cite{sterz2024quest,laux2024institutionalised}. 
The second response is epistemic: when role duality is unavoidable, system design should actively counteract the biases it introduces---for instance through mechanisms that surface disagreement between the HITL's operational judgment and independent benchmarks, thereby disrupting automation bias~\cite{laux2025automation}. Neither response eliminates the tension, but together they ensure that role duality is treated as a design problem to be mitigated rather than merely a conceptual observation to be acknowledged.

\section{Conclusion}

To operationalize human involvement in AI, we must move past the semantic noise. We propose a causal taxonomy: \textbf{HITL is constitutive}, making a human a necessary link in the chain; \textbf{HOTL is corrective}, placing a human in an external supervisory role with capacity to prevent, modify, or override outputs. We further distinguish temporal gradations within HOTL (synchronous, asynchronous, anticipatory) and note the interaction between causal roles and cognitive integration (complementary vs. hybrid intelligence).

Crucially, we posit that legal and ethical \enquote{Human Oversight} is structurally bound to the HOTL modality but adds normative requirements that go beyond mere causal positioning. The descriptive taxonomy we offer serves normative purposes: clear categories are a precondition for assessing whether systems meet regulatory requirements, since the criteria for effective oversight differ depending on whether a human occupies a HITL or HOTL configuration.

By adopting these distinctions, we can facilitate clearer communication between computer science, law, philosophy, psychology, and sociology, moving from metaphorical loops to concrete architectural and regulatory specifications.

\begin{credits}
\subsubsection{\ackname} Kevin Baum was supported by the German Research Foundation (DFG) under grant No. 389792660, as part of TRR 248, see https://perspicuous-computing.science, by the German Federal Ministry of Education and Research (BMBF) as part of the project MAC-MERLin (Grant Agreement No. 16IW24007), and by the European Regional Development Fund (ERDF) and the Saarland within the scope of (To)CERTAIN (Project ID EFRE-AuF-0000942). Johann Laux was supported by a British Academy Postdoctoral Fellowship (grant no. PF22\textbackslash220076) and the article is a deliverable of the \enquote{Emerging Laws of Oversight} project.

\subsubsection{\discintname} The authors have no competing interests.
\end{credits}

\bibliographystyle{splncs04}
\bibliography{bib}

\end{document}